\documentstyle[epsfig,pictex,preprint,aps]{revtex}
\tightenlines
\begin{document}
\title{\Large\bf Cosmological Light Element Bound 
on R Parity Violating Parameters}
\author{Jihn E. Kim$^{(a,b)}$, Bumseok Kyae$^{(a)}$ 
and Jong Dae Park$^{(a)}$} 
\address{$^{(a)}$Department of Physics and Center for Theoretical
Physics,\\ Seoul National University,
Seoul 151-742, Korea, and\\
$^{(b)}$School of Physics, Korea Institute for Advanced Study, 
Cheongryangri-dong, Dongdaemun-ku, Seoul 130-012, 
Korea}
\maketitle

\begin{abstract} 
In R parity violating theories, there does not exist the 
{\it stable} lightest
supersymmetric particle (LSP). The LSP, $\chi$, 
of MSSM decays to lighter ordinary particles with
dimension 6 effective R violating
interaction terms. Since the lifetime of $\chi$
can be sufficiently long for small R-violating couplings, 
it can affect the standard nucleosynthesis
scenario. This constraint gives an upper limit for
the lifetime of $\chi$ in the small $\tau_\chi$ region;  
$\sim\tau_\chi>10^6$ s. This translates to the sum of
squared R-violating couplings, for
$m_\chi=$ 60 GeV in the photino (bino) limit,
$(0.12 (0.05)|\lambda_{i,j,k}|^2+0.31(0.07)
|\lambda^\prime_{i,j\ne 3,k}|^2
+0.04(0.04)|\lambda^{\prime}_{i,3,k}|^2
+0.23(0.12)|\lambda_{i,j,k}^{\prime\prime}(i<j,k\ne 3)|^2)> 
7.7\times 10^{-24}$.
\end{abstract}

\newpage
\section{Introduction}

Supersymmetric extension of the standard model is beautiful
and probably needed toward a solution of the gauge hierarchy
problem \cite{review}. However, it introduces three outstanding 
problems: (i) the R-parity violating superpotential terms 
\cite{hall}, (ii) the $\mu$ problem \cite{mu}, and (iii) the 
flavor changing neutral currents \cite{dn}. There exist many
ideas to remedy these problems. 

In this paper, we concentrate on the R-parity violation and 
its probable effects in cosmology.
The problem in the minimal supersymmetric standard model is
that it allows the trilinear superpotential terms.
For decay processes, let us consider  
the trilinear R-violating couplings defined by 
\begin{equation}
W={1\over 2}\lambda_{ijk}L^i L^j \bar E^k
+\lambda^\prime_{ijk}L^i Q^j \bar D^k
+{1\over 2}\lambda^{\prime\prime}_{ijk}\bar D^i \bar D^j \bar U^k 
\end{equation}
where the superfields are denoted as the upper case Roman letters.
There can be the bilinear R-violating couplings, $\mu_i L_iH_2$,
which can be rotated away at the superpotential
level by redefining $H_1$ and $L_i$ fields. But if these bilinear
parameters are given at the electroweak scale with soft terms
present, one should worry about the correct minimum 
for the electroweak vacuum, implying that $\mu_i$ are physical
parameters \cite{mui}. But this subtle point is not of relevance
in our study of the neutralino decay.  
If both $\lambda^\prime$ and $\lambda^{\prime\prime}$ 
terms are present, the stability of proton is a function 
of $\lambda^\prime \lambda^{\prime\prime}$. To remove one of
these terms for proton stability, 
some employ discrete symmetries \cite{ir}. 
But the most widely employed discrete symmetry is the R-parity,
defined as $R=(-1)^{3B-L+2S}$ where B is the baryon number, L is
the lepton number and S is the spin of the particle. The fields
in the standard model have R = 1 while their superpartners have
R = --1. Namely superfields in Eq. (1) carries R charge --1, and 
all terms in Eq. (1) are forbidden by the R-parity conservation
assumption. In addition, if R-parity is conserved, there exists
an absolutely stable LSP (lowest mass supersymmetric particle) which
carries R = --1 and cannot decay to ordinary particles
due to the R conservation. This stable 
LSP can be a candidate for the constituent of the missing mass of
the universe.

The reason one considers the terms given in Eq. (1) is that they
are {\it not} forbidden by the gauge group and supersymmetry
alone. Therefore, phenomenologically
it is important to know bounds on the strengths of
the coupling parameters $\lambda,\lambda^\prime$ and $\lambda^{\prime
\prime}$. The most stringent laboratory bounds on these 
R-violating couplings are summarized in 
Table 1 \cite{bound}. However, a more severe bound comes from 
the lower limit of the lifetime of proton as mentioned
above. If we take the universal
strength for the couplings, then $\lambda$'s must be less than
$10^{-12}$ from the proton stability condition.  
With this strong constraint the search for new effects
in accelerator experiments may be futile. 

\vskip 0.5cm
\centerline{Table 1. The best phenomenological limits on R violating 
couplings for $M_{\tilde f}=100$ GeV \cite{bound}.}
\begin{center}
\begin{tabular}{|c|c|c|}
\hline
$\lambda_{ijk}$ &  $\lambda^\prime_{ijk}$ 
& $\lambda^{\prime\prime}_{ijk}$   \\
\hline
$(121)<0.05$ &  $(111)<0.001$ 
& $(112)< 10^{-6}$ \\
$(122)<0.05$ &  $(112)<0.02$   
&  $(113)< 10^{-5}$\\
\hline
\end{tabular}
\end{center}
\vskip 0.5cm

From the Table, we note that there exist several regions of 
the couplings allowed by the laboratory experiments, for
$\lambda,\lambda^\prime$ and $\lambda^{\prime\prime}$.
Generally, particle phenomenology gives an upper bound on
the couplings in view of nonobservation of rare processes.

On the other hand, cosmological constraints give different kinds
of bounds on the R-violating couplings. Some of the allowed
regions are reviewed in Ref.~\cite{sarkar}.
In this paper, we study the deuterium dissociation after nucleosynthesis,
and present a lower bound on the sum of squares of couplings.
There can be a window for the
allowed strength of the couplings. We try to estimate the 
lower bound of the couplings from deuterium dissociation 
as best as we can, including every possible
factor in the formulae. In fact, this topic was touched upon
earlier \cite{bouquet}, and there are a few papers regarding which
process (D or HE$^4$ dissociation) in the gravitino
decay gives a better bound \cite{pro}.   
In this paper, the lower bound on 
the R-violating couplings from the study of neutralino decay 
is expressed as an upper bound on the lifetime of the LSP ($\equiv \chi$).
Of course, if the lifetime is sufficiently long then our
analysis should not give any deuterium destruction since
deuteriums are too much separated.  This will happen sometime
after photon recombination. However, the study shows that
the upper limit of the lifetime falls in $10^6$ s region, and
we are safe from the recombination phenomena. But if the lifetime
is somewhat longer than the age of universe then the LSP has not
decayed yet, then the only constraint is
its mass density not exceeding the energy density
of the universe. This gives another
region of coupling constants, given as an upper bound, which is 
not considered here.

In Sec. 2, the decay rate of the neutralino is calculated. 
In Sec. 3, we calculate the decoupling
temperature and obtain $T_D\sim $ O(1) GeV.
In Sec. 4, these results 
are used to estimate the energy density of the neutralino and
deuterium dissociation effect from photons arising in 
the neutrlino decay. This study gives a bound on the sum of squares of
the R-violating couplings. Sec. 5 is a conclusion.

\section{Decay Rate}

With the superpotential given in Eq. (1), the LSP of MSSM $\chi$,
decays to ordinary particles. For example, with the above R-parity 
violating terms, one can write, for instance, a dimension 6 effective 
Lagrangian 
$$ 
-{\lambda^\prime_{ijk}\over M^2_{\tilde f}}(gN_2+{1\over 3}g_1N_1)
\int d^4\theta U^{aj}\chi\int d^2\theta E^i\bar D^k_a +{\rm h.c.}
$$
where $a$ is the color index, and 
$N_1$ and $N_2$ are defined by
\begin{equation}
\tilde \chi=N_1\tilde B+N_2\tilde W+N_3\tilde H_1^0+N_4\tilde H_2^0.
\end{equation}
For photino, one may
write 
$$
\chi=\tilde\gamma+F^{\mu\nu}\sigma_{\mu\nu}\theta+\cdots.
$$
But we use the vector multiplet notation in the Wess-Zumino
gauge
\begin{equation}
-\theta\sigma^\mu\bar\theta A_\mu(x)+i\theta\theta\bar\theta
\bar{\tilde\gamma}(x)
-i\bar\theta\bar\theta\theta{\tilde\gamma}(x)
+{1\over 2}\theta\theta\bar\theta\bar\theta D(x).
\end{equation}
For the decay
$\chi\rightarrow \bar u+d+e^+$, there exist 9 diagrams among which
one superfield diagram is
shown in Fig. 1. In total, there exist 1,080 (1,350) decay modes for
the decay of $\chi$ to any final states, excluding (including) 
the final state $t\bar t$ pair.

Assuming that $\chi$ is much heavier than b-quark, but lighter
than t-quark, and a universal sfermion mass, 
we obtain the decay width of $\chi$ as\footnote{For a formal expression
of matrix element squared, see \cite{baltz}.}
\begin{eqnarray}
&\Gamma_{\rm tot}={g^2\over 256\pi^3}{m^5_{\tilde\chi}\over
M^4_{\tilde f}}\Big[
(\sum|\lambda_{ijk}|^2)\{{1\over 8}N_2^2+{3\over 8}N_1^2\tan^2\theta_w\}
\nonumber\\
&+(\sum_{j\ne 3}|\lambda^\prime_{ijk}|^2)
\{{3\over 4}N_2^2+{7\over 12}N_1^2\tan^2\theta_w\}\nonumber\\
&+(\sum_{ik}|\lambda^\prime_{i3k}|^2)\{{3\over 8}N_2^2
+{7\over 24}N_1^2\tan^2\theta_w -{1\over 2}N_1N_2
\tan\theta_w\}\\
& +(\sum_{k\ne 3}|\lambda^{\prime\prime_{ijk}}|^2)\{
N_1^2\tan^2\theta_w\}
\Big].\nonumber
\end{eqnarray}
For $M_{\tilde f}=1$ TeV and $m_{\chi}=30$
GeV,\footnote{This number is a practical lower bound for
a neutralino mass in MSSM.} 
the numerical factor in front of the square bracket is
$1.8\times 10^{15}$ s$^{-1}$. If the largest value of the R-violating
coupling is of order $3\times 10^{-8}$, then the LSP lifetime falls in the
seconds region, which may affect the standard scenario of 
nucleosynthesis. 

\section{Decoupling Temperature}

The effect of a late decaying unstable (light) neutrino has been studied
two decades ago \cite{dicus}. Several years later,
the cosmological effect of a late decaying heavy particle 
has attracted a great deal of 
attention, due to the possible existence of 100 GeV scale
gravitino \cite{ellis}. The gravitino lifetime can be very long due
to the Planck mass suppression of the coupling constant. The lifetime
of the LSP in R-violating theories can be very long due to the smalless
of the R-violating parameters and three body final states. But their
roles in cosmology, being a late decaying particle, are similar.  

The decoupling temperature of the gravitino is close to the Planck
scale, so the inflation might have diluted them almost completely.
But after reheating, the gravitinos can be produced. Even though
the number density of these thermally produced gravitinos are
negligible compared to the number density of photon, its
cosmological effect is dramatic since the gravitino decay injects
huge energy in the cosmic soup \cite{ellis}. On the other hand,
the decoupling temperature $T_D$ of LSP is at a few GeV 
scale as shown below. 
If the reheating temperature after inflation is greater than the LSP
decoupling temperature $T_D$, then the LSP number density is
enormous compared to the thermally produced gravitino 
number density. This consideration will give a lower bound on the
R-violating coupling strengths. Of course, there exists another
region of allowed R-violating couplings, the upper bound of couplings, 
so that the LSP can be practically
considered as a stable particle in cosmology. 

The decoupling temperature of $\chi$, $T_D$, is determined by a competition
to the contribution to the number density of $\chi$ from
the expansion rate of the universe, i.e. the depletion of $\chi$
through $H\cdot$(the $\chi$ number density), 
and from the $\chi$ destruction rate, $\sigma\cdot$(relative velocity)
$\cdot$(useful number densities)$^2$. The destruction rate is
a function of sfermion mass for $\chi\chi\rightarrow f\bar f$. 
However, if all sfermion masses are much greater
than the $Z$ boson mass and if $\chi$ has a significant Higgsino
component, then the $Z$ exchange diagram can dominate. 
Since most superpartner masses are unknown at present, 
we illustrate for the latter possibility 
that the LSP has significant Higgsino components\footnote{For a more
concrete discussion, see, Ref. \cite{yama} 
}, i.e. $N_{3,4}$ of Eq. (3) are relatively large
$|N_{3,4}/\sqrt{N_1^2+N_2^2}|>
M_Z^2/M^2_{\tilde f}$, the production of $\chi$ is
dominated by the intermediate $Z$ boson diagram, and hence the
destruction cross section for $T_D$ estimate is given by
\begin{equation}
\sigma(\chi\chi\rightarrow e^+e^-)={G_F^2s\beta\over 24\pi^2}
(1-4x_w+8x^2_w)\left|{M_Z^2\over s-M_Z^2+iM_Z\Gamma_Z}\right|^2
(N_3^2-N_4^2)^2
\end{equation}
where $s$ is the total energy squared $x_w=\sin^2\theta_w\sim 0.232$ 
and $\beta$ is the velocity of $\chi$, 
$\beta=v/c=\sqrt{1-4m_\chi^2/s}$. We are interested in the
mass region $m_\chi<100$ GeV. 
For $2m_b<m_\chi<M_Z$, the channels $\chi\chi\rightarrow e^+e^-,
\mu^+\mu^-,\tau^+ \tau^-, \nu_e \bar \nu_e, \nu_\mu \bar\nu_\mu,
\nu_\tau \bar\nu_\tau, u\bar u, d \bar d, s\bar s, c\bar c$,
and $b\bar b$ are open. In this region, we must consider all
these final states, and Eq. (5) is changed as
$$
(1-4x_w+8x_w^2)\rightarrow (21-40x_w+{160\over 3}x_w^2)\simeq 14.6.
$$ 
For a given temperature, the Boltzmann
distribution of a massive particle is folded to the above
cross section. In the literature \cite{gondolo}, the thermal
average regarding the relative velocity of neutralino
has been properly taken into account. Thus, the Hubble expansion rate
and the reaction rate are related to find
the decoupling temperature, $<\sigma v_{12}>n_\chi\sim H$
where $v_{12}=|v_1-v_2|$ is the relative velocity,
\begin{eqnarray}
& {G_F^2m_\chi T\over 2\pi^2}
(21-40x_W+{160\over 3}x_w^2){M^4_Z\over (4m_\chi^2-M_Z^2)^2
+M_Z^2\Gamma^2_Z}\cdot\nonumber\\
&\cdot (N_3^2-N_4^2)^2
\cdot g \left({m_\chi T\over 2\pi}\right)^{3/2}e^{-{m_\chi\over T}}
=N_F^{1/2} {T^2\over M_{Pl}}\sqrt{8\pi^3\over 45}. 
\end{eqnarray}
We can take $g=2$ and $N_F=494/8$ in the region of our interest.
Then Eq. (6) gives a decoupling temperature through
\begin{equation}
{m_\chi\over T_D}\simeq \ln\left(
{G_F^2m_\chi^{5/2}T_D^{1/2}M_{Pl}g(21-40x_w+{160\over 3}x_w^2)M_Z^4(
N_3^2-N_4^2)^2\over \sqrt{N_F}
\sqrt{8\pi^3/45}(2\pi)^{5/2}[(4m_\chi^2-M_Z^2)^2+M_Z^2\Gamma_Z^2]}
\right)  
\end{equation}
For $m_\chi=30$ GeV and $|N_3^2-N_4^2|
\sim 0.1$, we obtain $T_D\sim 1.3$ GeV. In Table 2--4, we list the
decoupling temperature factor $d$, which is defined as
\begin{equation}
d={m_\chi\over T_D},
\end{equation} 
as a function of the neutralino mass, not only
for the Higgsino dominated cases \cite{review,dn} and also for the gaugino
dominated cases \cite{review,hall}. 
In these Tables, 
$M_{\tilde f}=300, 600$ and 1,000 GeV are considered,
and we distinguish four cases: [i] $N_1=0.99, N_2=0.14, N_3= N_4=0,$ 
[ii] $N_1=0.9, N_2=0.44, N_3=N_4=0$, [iii] $N_1=0.5, N_2=0.2, 
N_3=0.84, N_4=0$, [iv] $N_1=0.2, N_2=0.05, N_3=0, N_4=0.97$.
As one can see, the Higgsino
dominated cases give lower decoupling temperatures than the gaugino
dominated cases. This is because for most parameter space the Higgsino
dominated cases give larger cross sections than gaugino dominated
cases. This implies that the Higgsino dominated cases give smaller
relic densities than the gaugino dominated cases. For the $\chi$
dark matter possibility, $\chi$ may be dominated by 
gaugino.

If $|N_{3,4}|\ll 1$ so that $\chi$ is mostly
gaugino, then the $t$ and $u$ channel 
sfermion exchange diagrams are more important ones
for establishing equilibrium. This happens when $N_{1,2}
<M_Z^2/M^2_{\tilde f}\sim 0.1$ if $M_{\tilde f}>300$ GeV.
The mass dependence of the cross section times the
relative velocity is roughly $0.03m^2_\chi/M^4_{
\tilde f}$ and the decoupling temperature is $\sim$ 1.5 GeV
for the bino dominated case. For
the detailed result, see Table 2. We also note that the neutralino
decoupling temperature is insensitive to $\tan\beta$. In this
calculation, we used the expression given in Drees and Nojiri
\cite{yama}, which employed the partial wave analysis. In this
estimate, we set all the sfermion masses equal, for simplicity.
We also assumed 
\begin{equation}
{m_{A,h_1,h_2}\Gamma_{A,h_1,h_2}\over m^2_\chi}\simeq {1\over 1,500},\ \ 
{m_{A,h_1,h_2}\over m_\chi}\simeq {1\over 3}
\end{equation}
where $m_{A,h_1,h_2}$ are the masses of the pseudoscalar, the lighter
scalar and the heavier scalar, respectively, in the Higgs sector,
$\Gamma_{A,h_1,h_2}$ are their decay rates. But it turns out that
the cross sections and the decoupling temperatures are generally
insensitive to these parameters. In the calculation we also
take the limit $m_f/m_\chi=0$ where $m_f$ is the final fermion mass.
We also disregard the mixing of left and right sfermions because
these are not well known.

The gaugino and Higgsino components can decay to a fermion
and an anti-fermion through the $s$-channel Higgs boson
exchange of the three kinds ($A,h_1,h_2$). The Higgs boson
coupling to the fermion pair is proportional
to $m_d/M_W\cos\beta$ and $m_u/M_W\sin\beta$ for the
down-type and up-type quarks, respectively. So the Higgs boson
coupling is important only for the $t$ and $b$ (if $\tan\beta$ is very
large) quarks. Since we are interested in the light neutralino
case, we do not consider the follwing decay modes:
$\chi \chi\rightarrow t\bar t, W^+W^-, ZZ, Zh_{1,2}, ZA, W^{\pm}
H^{\mp}, h_{1,2}h_{1,2}, AA, h_{1,2}A$, and $H^+H^-$ where $H^{\pm}$
is the charged Higgs particle. 

If one consider the sfermion dominated 
region, $T_D$ falls in the O(1 GeV) region for the
interesting range of $M_{\tilde f}$ and $m_\chi$.\footnote{In view 
of the gauge hierarchy solution through supersymmetry, superpartner
masses are not far from the electroweak scale. Under this circumstance, 
the LSP mass might be below 100 GeV.} 
Then the number
density of neutralinos below the decoupling temperature is
about $\sim 10^{-9}$ times the photon number density, 
which is enormous. If it is stable,
neutralinos can form the missing mass of the universe.

\section{Lifetime Bounds}

If $\chi$ decays after 1 s, it causes problems during the nucleosynthesis
era. From Eq. (4), the lifetime $\tau_\chi$ is about $10^{-15}/
|\lambda|^2$ s. In the radiation dominated era, the cosmic time
is given by the cosmic temperature as
\begin{equation}
t=\left({90\over 32\pi^3N}\right)^{1/2}{M_{Pl}\over T^2}.
\end{equation}  
Long after the nucleosynthesis era, one constraint that the
neutralino energy density does not exceed the energy of relativistic
particles is 
\begin{equation}
m_\chi n_\chi(\tau_\chi)<(\pi^2/30)N_{\tau_\chi} T^4_{\tau_\chi}
\end{equation}
where $T_{\tau_\chi}$ is the temperature of the universe when
neutrlinos decay and $N_{\tau_\chi}$ is the effective degrees at the
time of $\chi$ decay. Using $N_{\tau_\chi}\sim 2.5$, since
the expected temperature at the decay time is less than $\sim$ keV, 
we obtain if $\tau_\chi$ is greater than 1 s,
\begin{equation}
{m_\chi\over T_{\tau_\chi}}<4.0\cdot 10^8
\end{equation}
where we used an approximate relation $T_D\simeq m_\chi/20$.
Using Eq. (6), we estimate
\begin{equation}
\tau_{\rm sec}<4.8\cdot 10^{10}\left({{\rm GeV}\over m_\chi}\right)^2.
\end{equation}
where $\tau_{\rm sec}$ is the lifetime of $\chi$ in units of seconds.
From Eq. (4), we estimate $\lambda$'s for the bino (photino)
case if the couplings are universal,  
\begin{equation}
|\lambda|>3.7 (6.4)\times 10^{-15}\left({m_\chi\over {\rm GeV}}\right).
\end{equation}

However, this is not the most stringent bound for $\tau_\chi>1$ s.
When $\chi$ decays long after the nucleosynthesis era,
the universe containing
light nuclei such as D, $^3$He, $^4$He and $^7$Li
might be affected. Since the 
observed abundances of these light nuclei are in good agreement 
with the conventional calculation of these elements through 
nucleosynthesis, we must ensure that the light ordinary particles
produced by $\chi$ decay should not dissociate these preciously
formed light nuclei.  

The most stringent bound comes from the study of the D dissociation
\cite{lin79,lin85}. The calculation of Ref. \cite{lin79} can be
expressed as
$$
(m_\chi f){\beta n_\chi\over E_* n_e}\le 1
$$
where $f=0.7$, $E_*=100$ MeV, $n_\chi/n_\gamma\sim 3\cdot 10^{-9}$,
and $\beta=0.23$\footnote{Depending on the allowed R-violating
couplings, $\beta$ can be different. Here, we take an eyeball number.
Anyway, a better estimate is given below.} 
at $T\simeq 100$ eV at the time of 
$\chi$ decay. If the above bound is not satisfied, $\chi$ should decay 
so that its number density changes drastically.
This implies that the parameters do not satisfy the above bound, then
the lifetime of $\chi$ must be less than 1 s, i.e. it decays before
D synthesis.  Assuming that $n_e\sim n_B$ (the baryon density), we
obtain for a given $\delta_B=n_B/n_\gamma$
$$
(m_\chi f)_{\rm GeV}< \{130\
({\rm for}\ \delta_B\sim 10^{-10}), 13\ ({\rm for}\ 
\delta_B\sim 10^{-9}), 1.3\
({\rm for}\ \delta_B\sim 10^{-8})\}.
$$
Namely, for interesting regions of the baryon asymmetry and the
$\chi$ mass the condition on the lifetime of $\chi$ requires a
detailed study.
 
However, this simple calculation \cite{lin79} does not include the effect
of $e^+e^-$ production from scattering on the background radiation
by the high energy photon debries from $\chi$ decay. This process
has been added for a better estimation of $n_e$ and the resulting
D dissociation rate \cite{lin85}. The Compton scattering and the
$e^+e^-$ production from scattering on the background radiation is
comparable at the critical temperature $T_*$ with the critical
photon energy $E_*$, satisfying
\begin{equation}
E_*T_*\simeq {1\over 50} ({\rm MeV})^2.
\end{equation}
At the cosmic temperature $T_*$, there are roughly $10^{-9}$ fraction
of photons having energy $w\ge 25T_*$ which can be used to produce
$e^+e^-$ pairs. Since these energetic photon number is comparable
to the electron number, the probability to scatter off the
electron through Compton scattering is comparable to the
probability to produce $e^+e^-$ pairs from background radiation,
and gives the relation (14).

When the photon energy exceeds $E_*$, the probability of D dissociation
is unimportant compared to the total cross section,
since the probability is proportional to
$\sigma_D/\sigma_{\rm tot}$ where $\sigma_D$ is the D dissociation
cross section and $\sigma_{\rm tot}$ is the total cross section. But the
total cross section decreases rapidly for the photon energy below
$E_*$. In this case, the $e^+e^-$ produced by the scattering on the
background radiation scatter off the low energy photons to transfer
the energy to photons. Thus, {\it there results some high energy photons 
with energy less than $E_*$ but high enough ($>2.225$ MeV) to 
dissociate D.} These photons are the major source of destructing D. 
In Ref. \cite{lin85} this effect has been taken into account.
It is summarized for the maximum allowable lifetime of $\chi$,
satisfying 
\begin{equation}
2x_0{\epsilon\over\epsilon_0}\int^{t_2}_{t_1}
\Sigma(\epsilon_0,T)e^{-t/\tau_{max}}{dt\over\tau_{max}}=1
\end{equation}
where $x_0$ is $n_\chi/n_e$ at the time of $\chi$ decay,
and $\epsilon$ is the energy of the electron.
We can take $t_1$ as the cosmic
time when the maximum secondary photon energy equals $E_*$
and $t_2$ as the cosmic time when the maximum scattered photon
energy equals the threshold energy for D dissociation. 
The above formula assumes a scaling behavior of $\Sigma$
\begin{equation}
\Sigma(\epsilon,T)={\epsilon\over\epsilon_0}\Sigma(\epsilon_0,T).
\end{equation}
The scaling behavior is not valid for low energy photons in which
case we use the original $\Sigma$,
\begin{equation}
\Sigma(\epsilon,T)=\int^{w_1}_{Q_D}{\sigma_D\over
\sigma_{KN}+\sigma_{pp}}2\left(1-{w\over w_m}\right){dw\over w_m}
\end{equation}
where $Q_D=2.225$ MeV. 
For low energy electrons the maximum scattered photon
energy is given by
\begin{equation}
w_m\simeq {12\gamma^2 T\over 1+{12\gamma T\over m_e}}\sim 12
\gamma^2 T.
\end{equation}
In Eq. (18), 
$\sigma_{KN}$ is the Klein-Nishina formula \cite{kn} and
$\sigma_{pp}$ is the pair production cross section off the proton.
For an accurate calculation, we will use Eq. (17). 

But for illustration, let us take the scaling law, 
Eq. (16), and calculate
$\Sigma$ at $E_*\sim 100$ MeV, in which case $\epsilon_0\sim 50$ MeV, 
and $T_*\sim 0.2$ keV. Then,
\begin{equation}
\Sigma(\epsilon_0,T)\sim {(w_1-Q_D)(2w_m-w_1-Q_D)\over w_m^2}
\left({\sigma_D\over \sigma_{KN}+\sigma_{pp}}
\right)_{E_\gamma=w_m,E_e=\epsilon_0}.
\end{equation}
In this case, $w_m\sim 24$ MeV, and Eq.(16) gives
\begin{equation}
\tau_{\rm max}={2.24\times 10^7\ {\rm s}\over
4.92+\ln M_{\rm GeV}-\ln\eta_9}
\end{equation}
where $M_{\rm GeV}=m_\chi/{\rm GeV}$ and $\delta_B=10^{-9}\eta_9$.
For $m_\chi=30$ GeV, the maximum allowable lifetime of $\chi$ is
$2.7\times 10^6$ s.

But in contrast to this simple illustration, the condition (16) is
rather complicated, and we must solve it numerically.
In Figures 2--9, we present the numerical solution of Eq. (16) 
for the maximum lifetime bounds of
the neutralino as a function of the neutralino mass. We remind again
that these figures are drawn with the assumption of the same
sfermion mass. But note that the almost Higgsino case does not
depend on the sfermion mass if it is much larger than the
$Z$ boson mass. The gaugino dominated case depends on the Higgsino 
mass, and here our assumption of the universal sfermion mass
is critical. Figures 2--4 correspond to $N_1=0.99,N_2=0.14,
N_3=N_4=0$, and Figures 5--7 correspond to $N_1=0.9, N_2=0.44,
N_3=N_4=0$. Fig. 8 and Fig. 9 correspond to $N_1=0.5, N_2=0.2, N_3=0.84,
N_4=0$, and $N_1=0.2,N_2=0,05,N_3=0,N_4=0.97$, respectively. 
Note that the $N_i$ dependence is meager, i.e. only
logarithmic.

For the gaugino dominated cases (Figs. 2--7), we translate this
bound to the lower bound on the sum of squared couplings
with appropriate coefficients. 
This condition gives a bound on the sum of
squares of R-violating couplings for a 60 GeV 
photino-like (bino-like) neutralino and for 
$\delta_B\simeq 5\times 10^{-10}$,
\begin{eqnarray}
&(\sum 0.12 (0.05)|\lambda_{i,j,k}|^2+\sum 0.31 (0.07)
|\lambda_{i,j\ne 3,k}^\prime|^2
+\sum 0.04(0.04)|\lambda^\prime_{i,3,k}|^2\nonumber\\
&+\sum 0.23(0.12)|\lambda
_{i,j,k}^{\prime\prime}(i<j,k\ne 3)|^2)>7.7\cdot 10^{-24}.
\end{eqnarray}
Thus a generic bound on the R-violating couplings, if 
the couplings are of the same order, is about $2.8\times 10^{-12}$,
which is a better bound than Eq. (14).

\section{Conclusion}

We have studied the R-parity violation effects on the
dissociation of light nuclei, in particular to deuterium.
We tried to include every factor in the formulae so
that the estimate is reliable within a factor of a few.
From this study, we obtain that the lifetime of the 
neutralino LSP must be shorter than $\sim 2\times 10^{6}$ s, which
translates to the sum of squared R-violating couplings 
being greater than $10^{-22}$. Therefore, the lifetime
of the neutralino between $2\times 10^6$ s and $\sim {\rm a\ few}
\times 10^{17}$ s is forbidden. 

If the strengths of R-violating couplings are comparable,
which is possible in some spontaneous breaking scenario
of R-violation, the lower bound in the large coupling region
is $\sim 10^{-12}$. Such a small coupling implies a very
small ratio for the mass parameters, 
$\tilde v/M_P\sim 10^{-12}$ where $\tilde v$ is the scale of 
spontaneous R-parity violation. Thus the lower 
limit of R-parity violating scale is far below the intermediate scale
for the gaugino condensation and the axion decay constant.   
But this region is almost forbidden by the proton stability
condition, which requires $\lambda^\prime\lambda^{\prime\prime}
<10^{-24}$. 
With the universal strength assumption on the R-violating
couplings and to keep
the $\lambda>10^{-12}$ to be viable for some region of
the coupling space, $\lambda^\prime$
or $\lambda^{\prime\prime}$ must be required to vanish,
e.g. from an additional symmetry.

\acknowledgments
This work is supported in part by KOSEF,  MOE through
BSRI 98-2468, and Korea Research Foundation.

\newpage
\noindent{\bf Table Captions}
\vskip 0.5cm
\noindent Table 2. The decoupling temperature factor $d$
for $M_{\tilde f}=300$ GeV. 
The decoupling temperature is given by $T_D=m_\chi/d$. $N_i$
are defined in $\chi=N_1\tilde B+N_2\tilde W+N_3\tilde h_1
+N_4\tilde h_2$. The four cases for $N_i$ are [i] $N_1=0.99, 
N_2=0.14, N_3= N_4=0,$ [ii] $N_1=0.9, N_2=0.44, N_3=N_4=0$, 
[iii] $N_1=0.5, N_2=0.2, N_3=0.84, N_4=0$,
and [iv] $N_1=0.2, N_2=0.05, N_3=0, N_4=0.97$.
\vskip 0.6cm

\noindent Table 3. The decoupling temperature factor $d$. 
Same as Table 2 except for $M_{\tilde f}=600$ GeV.
\vskip 0.6cm 

\noindent Table 4. The decoupling temperature factor $d$. 
Same as Table 2 except for $m_{\tilde f}=1,000$ GeV.

\newpage
\centerline{Table 2}
\vskip 0.5cm
\begin{center}
\begin{tabular}{|c|c|c|c|c|c|}
\hline
$m_\chi$ [GeV] &  $\tan\beta $ & [i] & [ii]  & [iii]  & [iv] \\
\hline
 30 & 2 & 20.25& 20.32& 32.83 & 33.40  \\
 30 & 10 & 20.25& 20.32& 32.83& 33.40  \\
 30 & 30 & 20.25& 20.32& 32.84& 33.40  \\
 30 & 50 & 20.25& 20.32& 32.85& 33.40  \\
 60 & 2 & 21.49& 21.56& 29.26& 29.83  \\
 60 & 10 & 21.49& 21.56& 29.27& 29.83  \\
 60 & 30 & 21.49& 21.56& 29.36& 29.83  \\
 60 & 50 & 21.49& 21.56& 29.52& 29.83  \\
 100 & 2 & 22.24& 22.31 & 29.49& 30.06   \\
 100 & 10 & 22.24& 22.31& 29.50& 30.06  \\
 100 & 30 & 22.24& 22.31& 29.54& 30.06  \\
 100 & 50 & 22.24& 22.31& 29.64 & 30.06  \\
 150 & 2 & 22.62& 22.69& 28.97& 29.53  \\
 150 & 10 & 22.62& 22.69& 28.97& 29.53  \\
 150 & 30 & 22.62& 22.69& 29.03& 29.53  \\
 150 & 50 & 22.62& 22.69& 29.17& 29.53  \\
\hline
\end{tabular}
\end{center}

\newpage
\centerline{Table 3.}
\vskip 0.5cm
\begin{center}
\begin{tabular}{|c|c|c|c|c|c|}
\hline
$m_\chi$ [GeV] &  $\tan\beta $ & [i] & [ii]  & [iii]  & [iv] \\
\hline
 30 & 2 & 17.58& 17.64 & 32.83& 33.40  \\
 30 & 10 & 17.58& 17.64& 32.83& 33.40  \\
 30 & 30 & 17.58& 17.64& 32.84& 33.40  \\
 30 & 50 & 17.58& 17.64& 32.85& 33.40  \\
 60 & 2 & 18.90& 18.97& 29.26& 29.83  \\
 60 & 10 & 18.90& 18.97& 29.27& 29.83  \\
 60 & 30 & 18.90& 18.97& 29.34& 29.83  \\
 60 & 50 & 18.90& 18.97& 29.47& 29.83  \\
 100 & 2 & 19.83& 19.90& 29.49 &  30.06 \\
 100 & 10 & 19.83& 19.90& 29.50& 30.06  \\
 100 & 30 & 19.83& 19.90& 29.52& 30.06  \\
 100 & 50 & 19.83& 19.90 & 29.58& 30.06  \\
 150 & 2 & 20.49& 20.56& 28.97&  29.53 \\
 150 & 10 & 20.49& 20.56& 28.97& 29.53  \\
 150 & 30 & 20.49& 20.56& 29.00& 29.53  \\
 150 & 50 & 20.49& 20.56& 29.06& 29.53  \\
\hline
\end{tabular}
\end{center}
\vskip 0.5cm

\newpage
\centerline{Table 4.}
\vskip 0.5cm
\begin{center}
\begin{tabular}{|c|c|c|c|c|c|}
\hline
$m_\chi$ [GeV] &  $\tan\beta $ & [i] & [ii]  & [iii]  & [iv] \\
\hline
 30 & 2 & 15.60& 15.67& 32.83&  33.40 \\
 30 & 10 & 15.60& 15.67& 32.83& 33.40  \\
 30 & 30 & 15.60& 15.67& 32.84& 33.40  \\
 30 & 50 & 15.60& 15.67& 32.86& 33.40   \\
 60 & 2 & 16.94& 17.00& 29.26&  29.83 \\
 60 & 10 & 16.94& 17.00& 29.27& 29.83  \\
 60 & 30 & 16.94& 17.00& 29.34& 29.83  \\
 60 & 50 & 16.94& 17.00& 29.47& 29.83  \\
 100 & 2 & 17.90& 17.97& 29.49&  30.06 \\
 100 & 10 & 17.90& 17.97& 29.50& 30.06  \\
 100 & 30 & 17.90& 17.97& 29.52& 30.06  \\
 100 & 50 & 17.90& 17.97 & 29.56& 30.06  \\
 150 & 2 & 18.65& 18.72& 28.97&  29.53 \\
 150 & 10 & 18.65& 18.72& 28.97& 29.53  \\
 150 & 30 & 18.65& 18.72& 28.97& 29.53  \\
 150 & 50 & 18.65& 18.72& 29.03& 29.53  \\
\hline
\end{tabular}
\end{center}
\vskip 0.5cm

\begin{figure}[htb]
$$\beginpicture
\setcoordinatesystem units <72pt,18pt> point at 0 0
\setplotarea x from -0.700 to 1.800, y from -1.000 to 2.100 
\setplotsymbol ({\normalsize.})
\setsolid
\setlinear
\plot
 -0.700  0.900  0.200  0.900  1.000  2.100 
/
\setlinear
\plot
 0.200  0.900  1.000  0.400  1.800  0.800
/
\plot
 1.000  0.400  1.800  -1.000   
/
\put{$\slash$} at 0.5 0.70
\put{$\chi$} at -0.5 1.2
\put{$Q$} at 1.10 2.1
\put{$L$} at 1.9 0.8
\put{$\bar D$} at 1.9 -1.0
\endpicture$$
\caption{A superfield diagram for the gaugino component 
of $\chi$. The intermediate state is $Q$. There exists another 
diagram with the external $Q$ and $L$ lines exchanged.}
\end{figure}

\newpage
\begin{figure}[ht]
\begin{center}
\epsfig{file=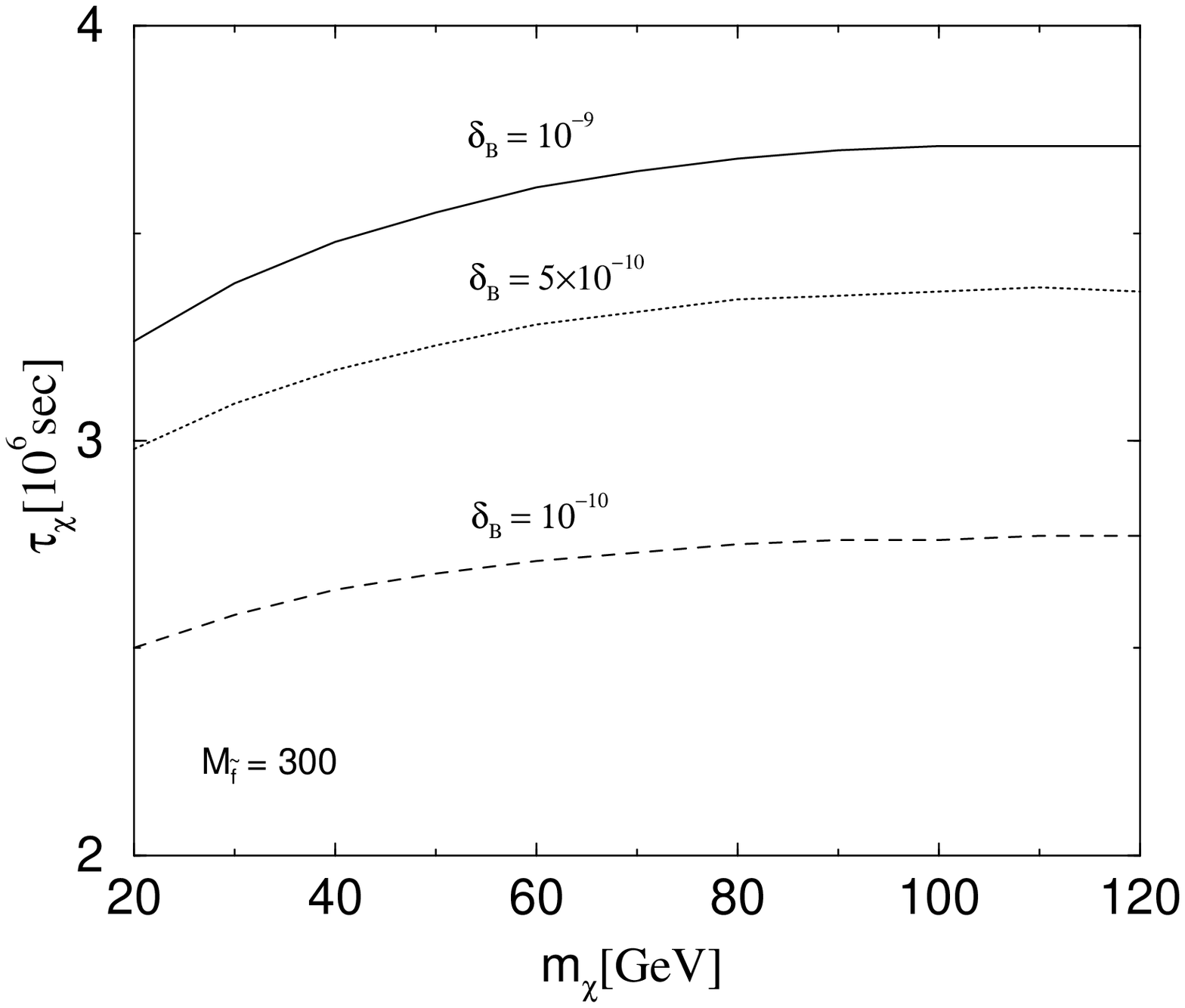, height=8 cm, width=10 cm}
\caption{The maximum allowable lifetime of $\chi$
for
$M_{\tilde f}=300$ GeV. $N_1=0.99,N_2=0.14,N_3=N_4=0.$}
\vspace{1 cm}
\epsfig{file=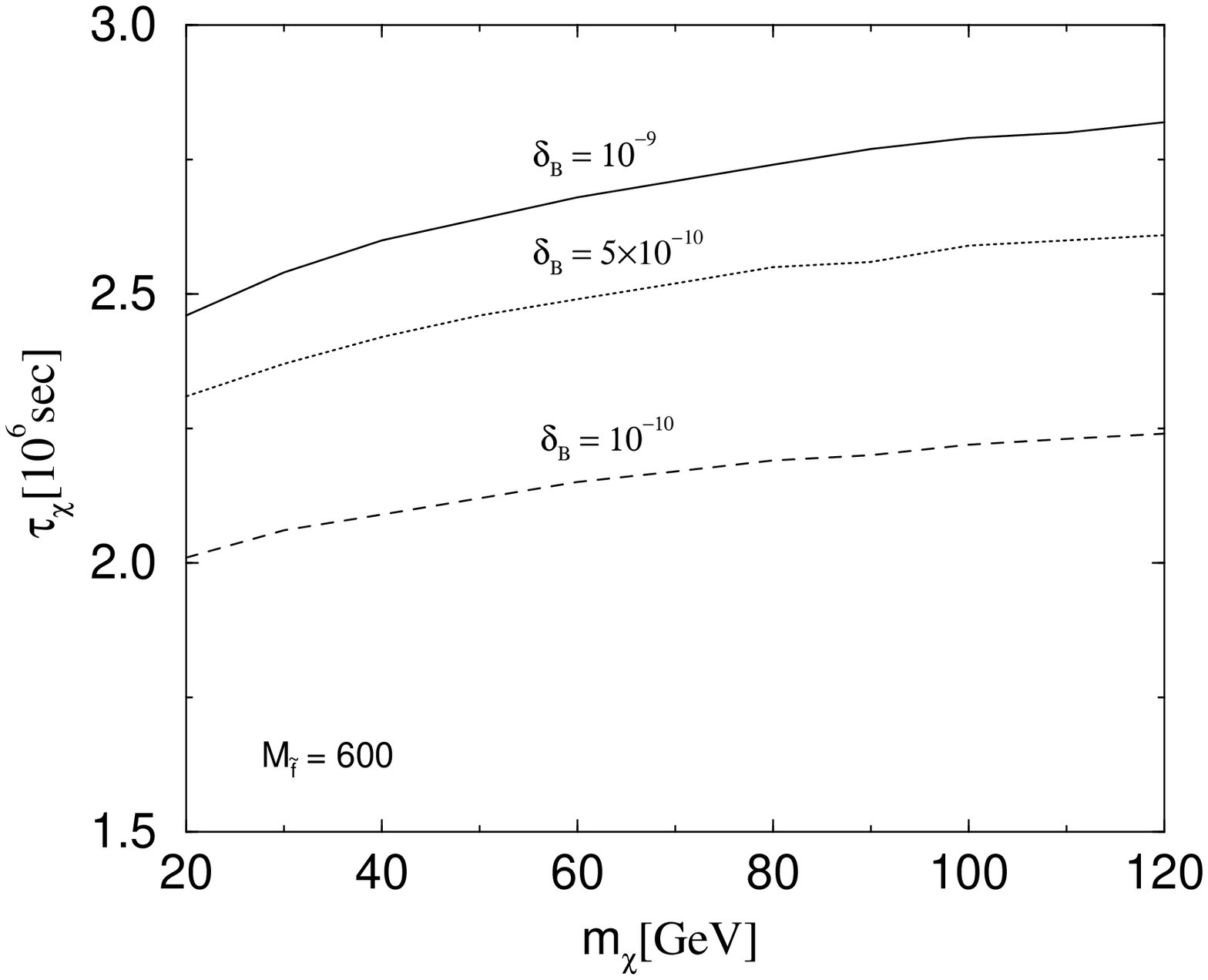, height=8 cm, width=10.5 cm}
\caption{Same as Fig. 2 except
for $M_{\tilde f}=600$ GeV.}
\end{center}
\end{figure}

\newpage
\begin{figure}[ht]
\begin{center}
\epsfig{file=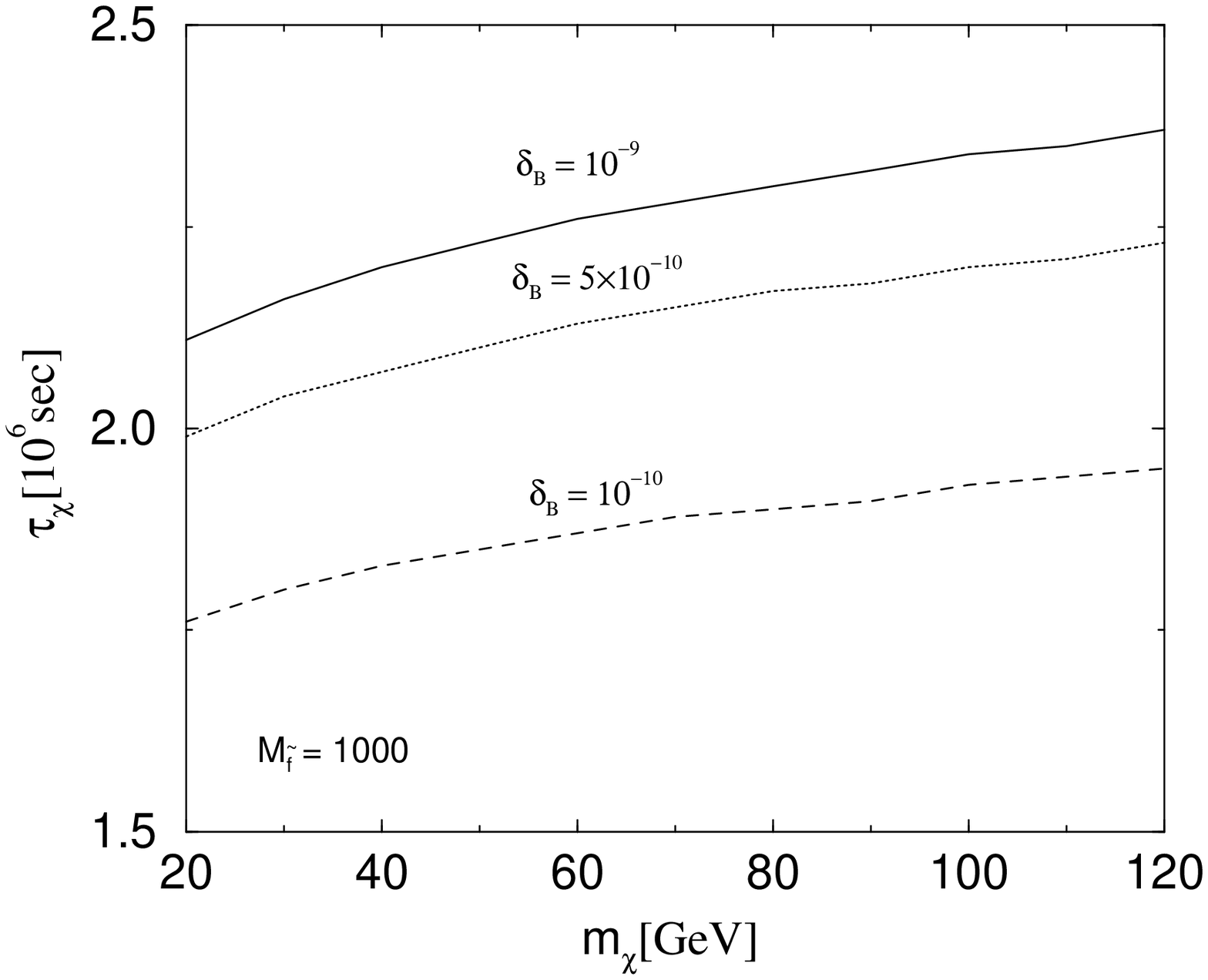, height=8 cm, width=10.5 cm}
\caption{Same as Fig. 2 except for $M_{\tilde f}=1,000$ GeV.}
\vspace{1 cm}
\epsfig{file=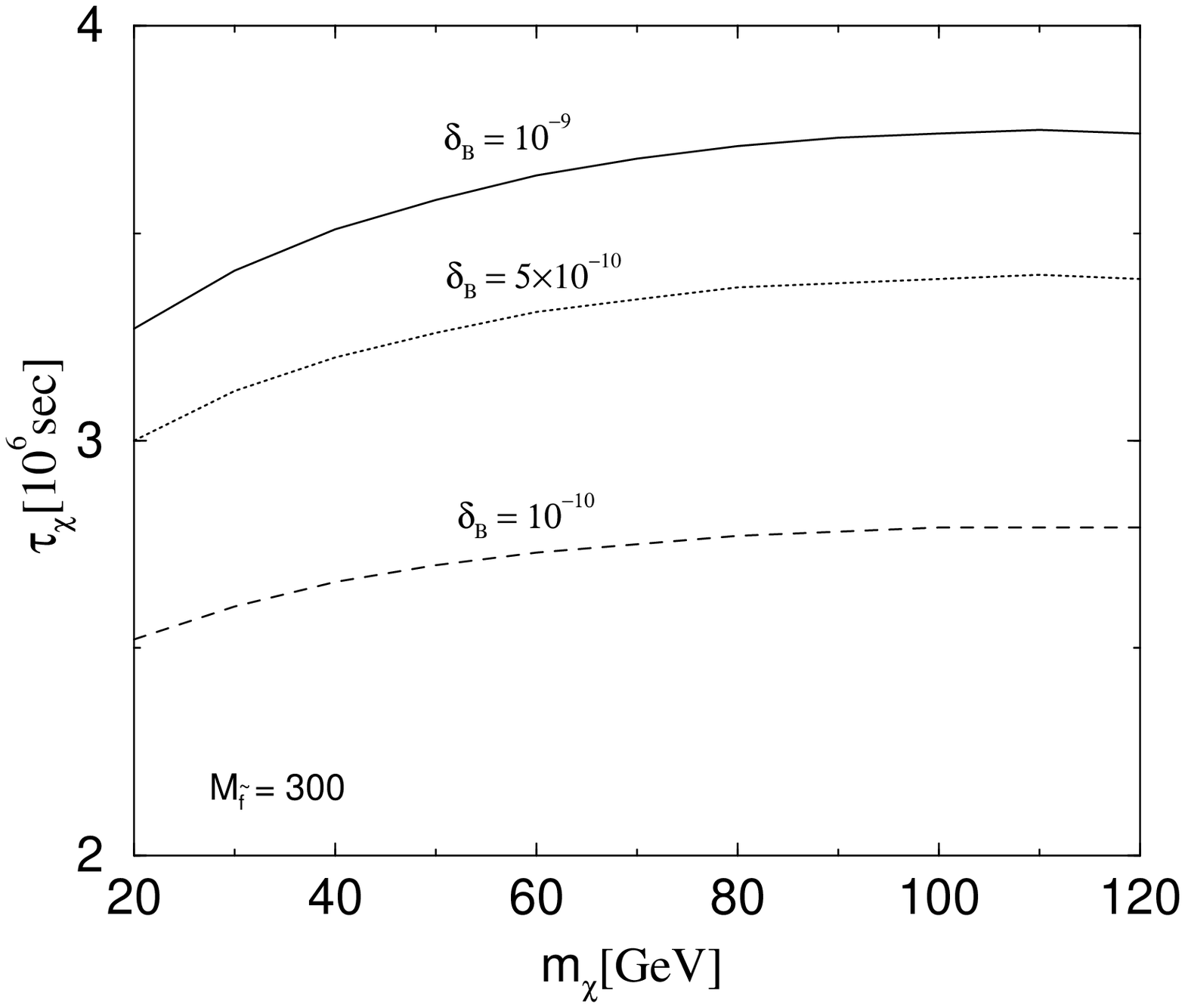, height=8 cm, width=10 cm}
\caption{Same as Fig. 2 except for
$N_1=0.9,N_2=0.44,N_3=N_4=0$
and $M_{\tilde f}=300$ GeV.}
\end{center}
\end{figure}

\newpage
\begin{figure}[ht]
\begin{center}
\epsfig{file=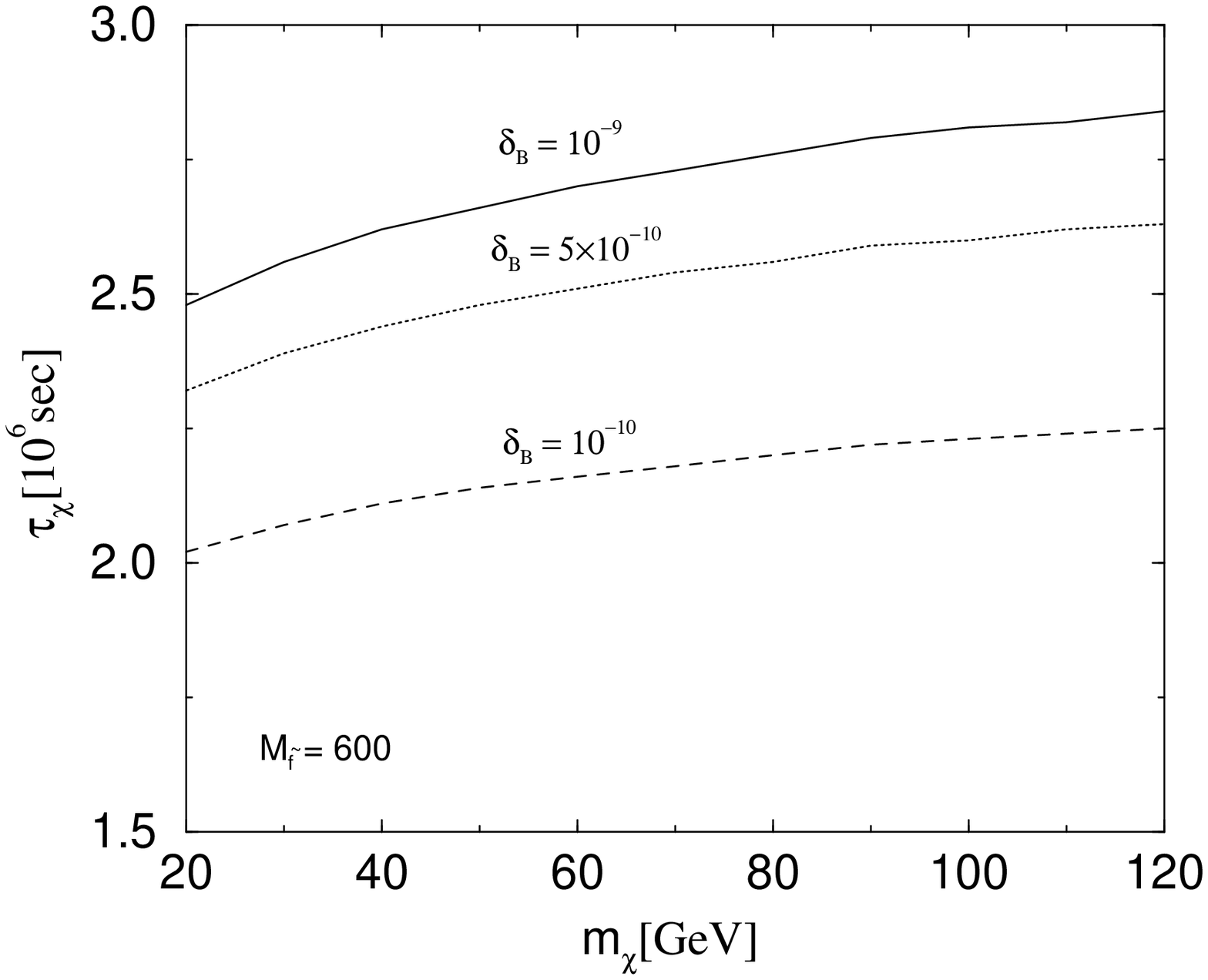, height=8 cm, width=10 cm}
\caption{Same as Fig. 5 except for $M_{\tilde f}=600$ GeV.}
\vspace{1 cm}
\epsfig{file=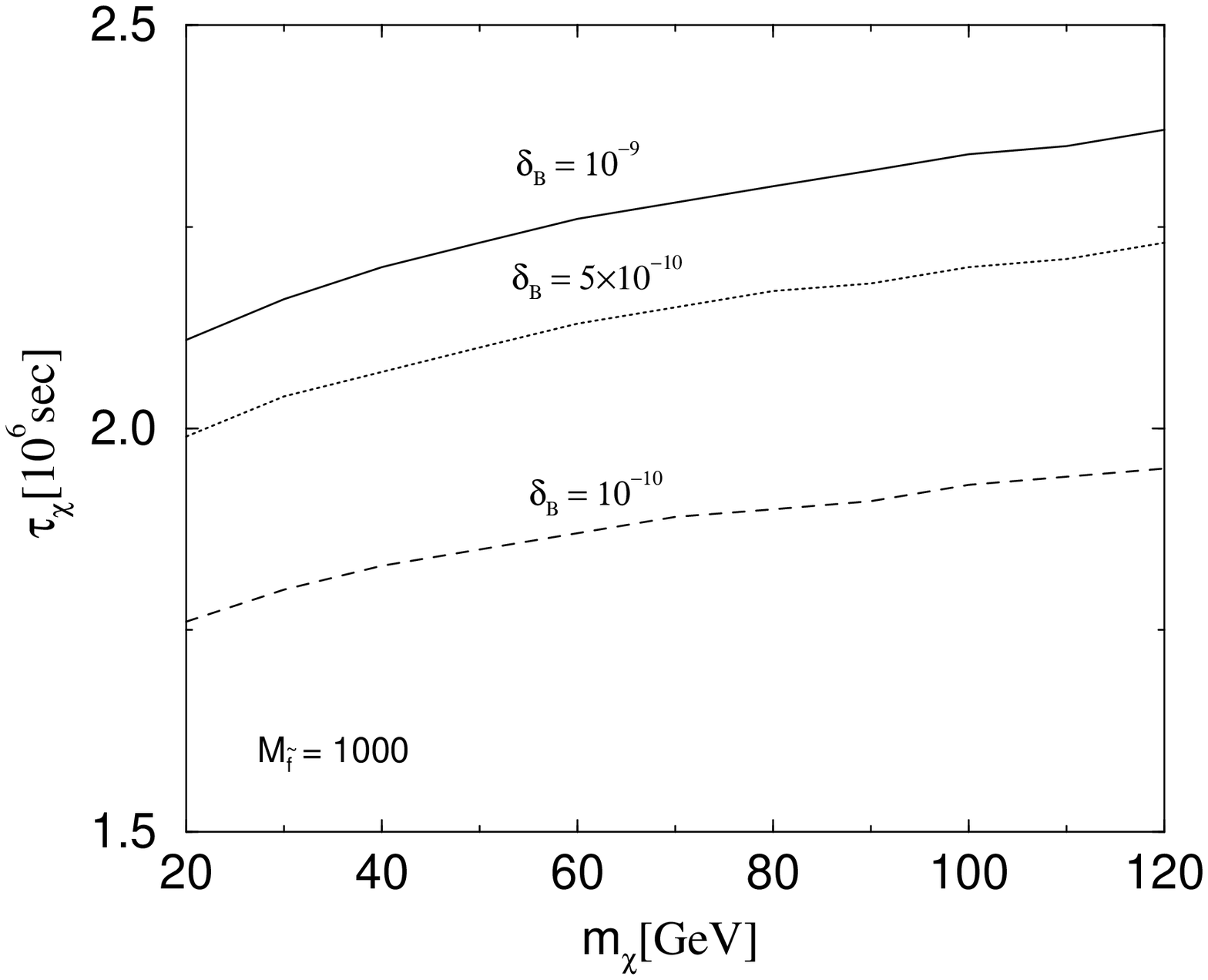, height=8 cm, width=10 cm}
\caption{Same as Fig. 5 except for $M_{\tilde f}=1,000$ GeV.}
\end{center}
\end{figure}

\newpage
\begin{figure}[ht]
\begin{center}
\epsfig{file=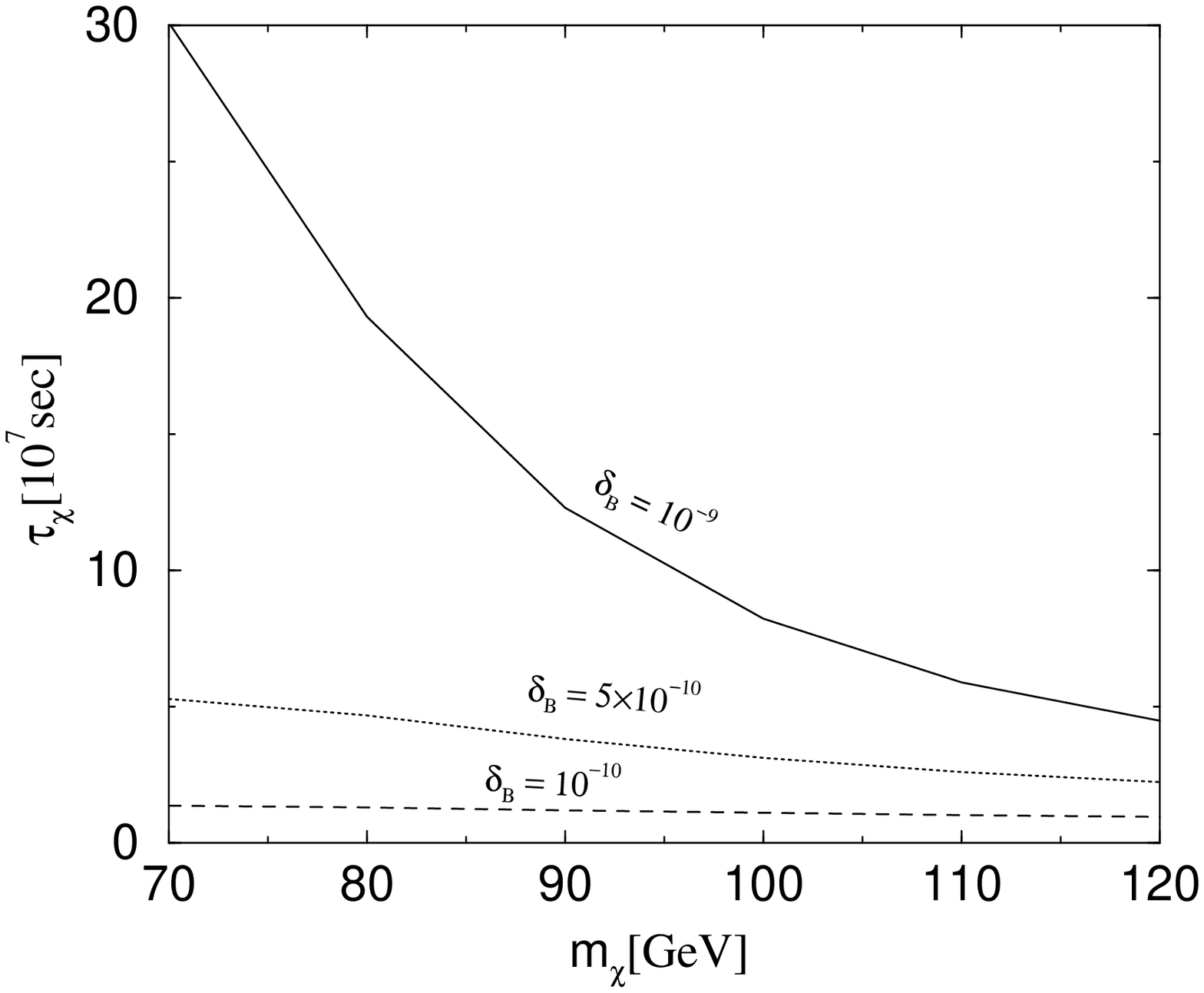, height=8 cm, width=10 cm}
\caption{The upper limit of $\tau_\chi$ for $N_1=0.5, N_2=0.2,
N_3=0.84, N_4=0$.
The $M_{\tilde f}$ dependence is negligible.}
\vspace{1 cm}
\epsfig{file=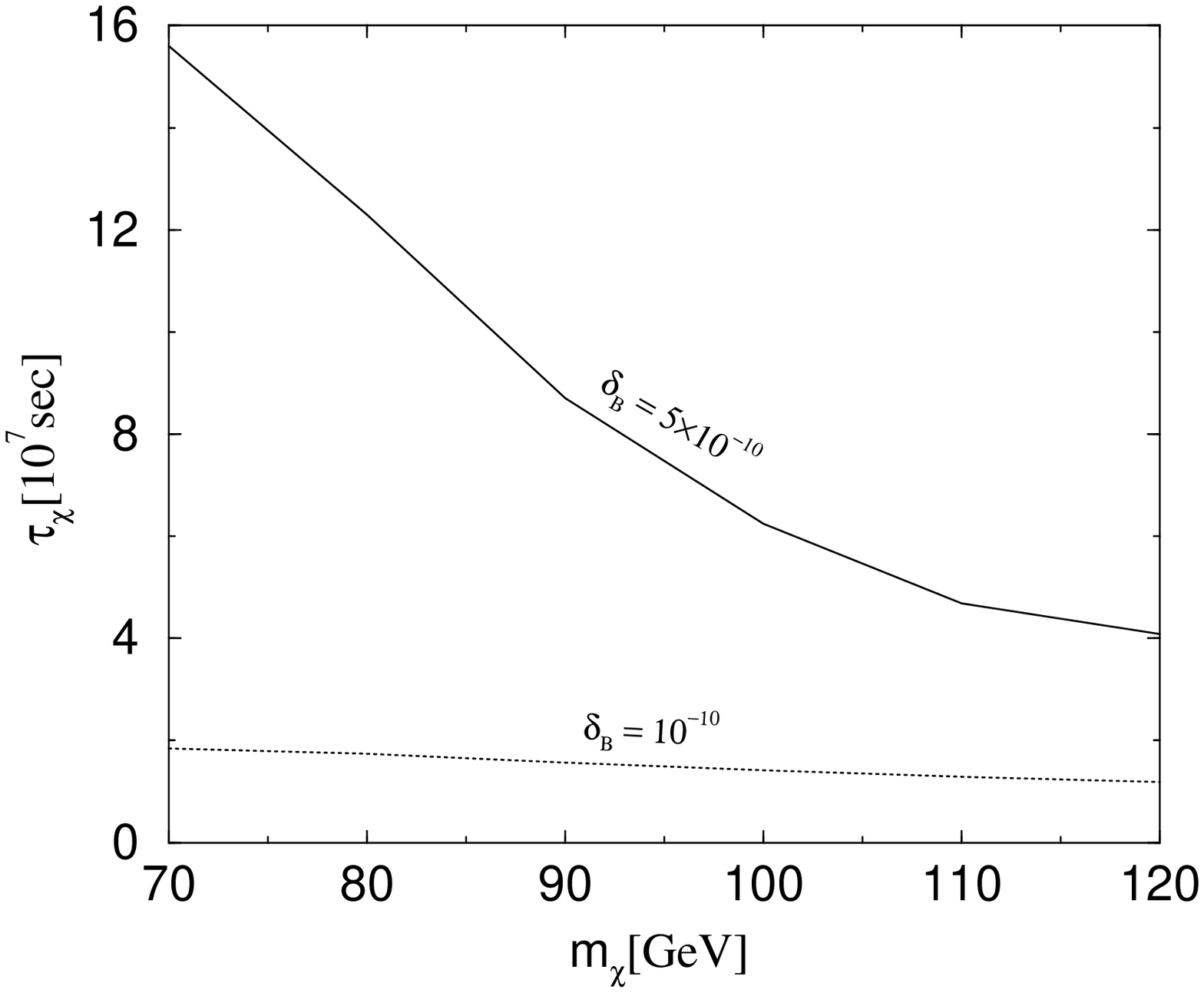, height=8 cm, width=10 cm}
\caption{Same as Fig. 8 except for $N_1=0.2, N_2=0.05,
N_3=0,N_4=0.97$.}
\end{center}
\end{figure}

\end{document}